\documentclass[10pt,conference]{IEEEtran}
\usepackage{graphicx}
\usepackage{float}
\usepackage{subcaption}
\usepackage{tabularx}
\usepackage{todonotes}
\usepackage{tcolorbox}
\usepackage{amsmath}

\usepackage{amssymb}
\usepackage{balance}
\usepackage{booktabs}
\usepackage{caption}
\usepackage{colortbl}
\usepackage{xcolor}
\usepackage{enumitem}
\usepackage{comment}
\usepackage{soul}
\usepackage{hyperref}

\title{A Blockchain-Oriented Software Engineering Architecture for Carbon Credit Certification Systems}

\author{
\IEEEauthorblockN{
Matteo Vaccargiu\IEEEauthorrefmark{1},
Azmat Ullah\IEEEauthorrefmark{2},
Pierluigi Gallo\IEEEauthorrefmark{3}
}
\IEEEauthorblockA{\IEEEauthorrefmark{1}
University of Cagliari, Italy\\
Email: matteo.vaccargiu@unica.it
}
\IEEEauthorblockA{\IEEEauthorrefmark{2}
University of Camerino, Italy\\
Email: azmat.ullah@unicam.it
}
\IEEEauthorblockA{\IEEEauthorrefmark{3}
University of Palermo, Italy\\
Email: pierluigi.gallo@unipa.it
}
}

\begin{document}

\maketitle

\begin{abstract}
Carbon credit systems have emerged as a policy tool to incentivize emission reductions and support the transition to clean energy. Reliable carbon-credit certification depends on mechanisms that connect actual, measured renewable-energy production to verifiable emission-reduction records. Although blockchain and IoT technologies have been applied to emission monitoring and trading, existing work offers limited support for certification processes, particularly for small and medium-scale renewable installations. This paper introduces a blockchain-based carbon-credit certification architecture, demonstrated through a 100 kWp photovoltaic case study, that integrates real-time IoT data collection, edge-level aggregation, and secure on-chain storage on a permissioned blockchain with smart contracts. Unlike approaches focused on trading mechanisms, the proposed system aligns with European legislation and voluntary carbon-market standards, clarifying the practical requirements and constraints that apply to photovoltaic operators. The resulting architecture provides a structured pathway for generating verifiable carbon-credit records and supporting third-party verification.
\end{abstract}


\section{Introduction}

Carbon credit systems have emerged as a primary tool to incentivise emission reductions and supporting the transition to clean energy. These mechanisms reward verified carbon-saving activities, such as renewable energy production or sustainable agriculture, with tradable certificates that entities can purchase to offset their own emissions. However, current credit generation and exchange systems suffer from inefficiency, high transaction costs, and lack of transparency \cite{en17194811,8486626}.

To address these limitations, recent research has proposed the combination of blockchain and Internet of Things (IoT) technologies to automate and secure credit certification and trading \cite{en17194811,liu2025blockchain}. IoT sensors can monitor emissions, energy use, and environmental parameters, while blockchain provides a tamper-proof, distributed ledger where credits can be issued, tracked, and retired using smart contracts. This combination has been shown to reduce transaction costs and improve auditability \cite{8486626,earth5030020}. 

The integration of IoT sensors with blockchain smart contracts lowers transaction costs through three main methods. Firstly, smart contracts remove the need for intermediaries such as brokers, verification agents, and registries that usually authenticate carbon credits, thereby cutting related fees \cite{8486626}. Secondly, the automated validation of continuous IoT data streams decreases the need for manual verification and reduces audit frequency \cite{en17194811}. Third, our data aggregation techniques cut blockchain expenses by batching minute-level readings into five-minute intervals, decreasing daily transactions from 1,440 to 288 and reducing operational costs by roughly 80\%.

In particular, Liu et al. \cite{liu2025blockchain} propose a layered system architecture for the Chinese automotive sector that highlights the need for standardized, traceable and policy-aligned infrastructure in national carbon-credit schemes.

The design and implementation of such integrated systems pose specific challenges for software engineering. As noted in recent surveys \cite{10.1145/3530813}, blockchain-based software requires specialized architectures, development processes, and testing strategies. Smart contracts, small programs that enforce agreement logic, are difficult to test and validate, especially when interfacing with real-world data and domain-specific logic \cite{8847638}. Developer engagement with sustainability concerns in blockchain projects 
has increased, though energy consumption and optimization challenges remain 
prominent topics requiring continued attention \cite{10.1145/3661167.3661194, 10621724}. While much attention has been given to the financial applications of smart contracts\cite{10123598}, their role in environmental certification requires additional considerations, including legal validity, audit requirements, and sustainability alignment.

This paper presents a software architecture to support the certification of carbon credits using blockchain and IoT. Our goal is to define a practical and legally informed software solution that can support the traceability and audit of credits generated by agricultural energy producers and similar other sectors.

The architecture consists of IoT sensors for energy data collection, smart contracts on Hyperledger Fabric for credit generation and registration, and a dashboard interface for operators. Credits are generated only when predefined environmental and legal criteria are met. Although similar ideas have been explored in previous conceptual platforms \cite{en17194811,liu2025blockchain}, our contribution provides a structured 
architectural design that formalizes the integration of IoT monitoring with blockchain certification and explicitly aligns the system with European regulatory requirements (EU ETS, Effort Sharing Regulation, EU carbon removals framework) \cite{directive2003ETS,reg2018effort,reg2024carbon} and voluntary carbon market standards. Furthermore, we ground our design in the socio-technical barriers identified in prior deployments, including device reliability, network accessibility, and the complexity of integrating heterogeneous data \cite{su152014843,earth5030020}.

The architecture supports requirements from voluntary carbon markets and sustainability reporting frameworks. Specifically, it aligns with the UN Sustainable Development Goals (SDG 7: Affordable and Clean Energy; SDG 13: Climate Action) \footnote{https://sdgs.un.org/goals} and enables Environmental, Social, and Governance (ESG) disclosures by providing auditable emission-reduction records. This offers researchers and practitioners infrastructure for certification systems that balance automation, compliance, and transparency.

The remainder of this paper is organized as follows: in Section~\ref{relwork} we present the related work, highlighting the novelty of our contribution; in Section~\ref{meth} we describe the proposed methodology for carbon credit certification; in Section~\ref{impl} we discuss implications for practitioners in light of current regulations and in Section~\ref{limit} potential limitation; finally, we conclude the paper in Section~\ref{conclusion}.

\section{Related Works}
\label{relwork}

Several studies have explored blockchain-based carbon credit trading systems. Boumaiza and Maher \cite{en17194811} propose a blockchain- and IoT-based platform for carbon-credit exchange, where IoT sensors monitor emissions and smart contracts automate credit issuance and trading among participants. The system focuses on transparency, fraud reduction, and real-time verification, enabling efficient digital carbon-credit exchange. Muzumdar et al. \cite{muzumdar2022permissioned} propose a Hyperledger-based emission trading system with a priority-based auction mechanism, achieving 84.37\% bid satisfaction with 0.41s transaction latency. Liu et al. \cite{liu2025blockchain} develop an RFID and blockchain infrastructure for China's automotive carbon-credit market using five types of smart contracts to manage credit lifecycles. Saraji and Borowczak \cite{saraji2021blockchainbasedcarboncreditecosystem} present a carbon credit ecosystem on Ethereum using automated market makers (AMM) for tokenized carbon credit trading, focusing on marketplace liquidity and price discovery mechanisms for existing credits.

Other works focus on emission monitoring and verification. Effah et al. \cite{effah2021carbon} implement a monitoring and trading system on FISCO BCOS that transmits emission data every 24 hours to smart contracts that automatically trigger trading when credits fall below thresholds. Woo et al. \cite{woo2021applying} review blockchain for building energy performance MRV, finding that systems like Hyperledger Fabric can provide digital MRV with reduced environmental impact versus Proof-of-Work algorithms. Basu et al. \cite{basu2023blockchain} propose a framework for carbon supply chain transparency across multi-institution networks, tracking emissions from component manufacturers to final assembly.

Technical implementation studies examine scalability and privacy. Westphall and Martina \cite{westphall2022blockchain} analyze Hyperledger Fabric scalability in energy trading, achieving maximum throughput with 5,000 sensors, buyers, and sellers, implementing pseudonymity through identity mixing and off-chain payment. Zhang et al. \cite{zhang2018blockchain} present early work on Hyperledger Framework for Internet of Energy, describing transaction processes with Fabric-CA, peer endorsement nodes, and orderer consensus nodes. Xu et al. \cite{xu2024blockchain} develop GPChain for construction industry CMP certification, implementing privacy-preserving homomorphic encryption for carbon footprint recording across supply chains. Oracle-based solutions have been proposed to bridge blockchain platforms with real-time energy data from smart meters, enabling automated peer-to-peer 
energy trading \cite{10731800}. Comprehensive assessment of blockchain's role in achieving Sustainable Development Goals demonstrates that energy sector applications provide benefits extending beyond technical aspects to include social, economic, and environmental dimensions \cite{su152014843}.

Existing works primarily address trading mechanisms \cite{en17194811,muzumdar2022permissioned,liu2025blockchain} or monitoring \cite{effah2021carbon,woo2021applying} with limited focus on certification processes. Our contribution differs by focusing on certification systems rather than trading, grounding the architecture in European regulatory requirements (EU ETS \cite{directive2003ETS}, Effort Sharing Regulation \cite{reg2018effort}, voluntary market standards \cite{battocletti2023voluntary}), and targeting small-scale renewable energy producers. Unlike systems assuming credits exist \cite{en17194811,muzumdar2022permissioned}, our four-layer architecture models complete data flow from energy measurement to blockchain recording for third-party verification. While Xu et al. \cite{xu2024blockchain} develop GPChain for certifying the carbon footprints of construction materials and products—focusing on privacy-preserving data sharing and access-controlled smart contracts—their system does not address carbon-credit issuance or alignment with voluntary market standards. In contrast, our design targets renewable-energy producers and models the full certification workflow required by EU and voluntary carbon-market regulations, spanning IoT-based energy measurement to blockchain verification.

\section{Methodology} \label{meth}

This section presents the architectural framework through a case study of a 100 kWp solar photovoltaic system. The PV case study demonstrates how IoT-based monitoring can be integrated with blockchain certification for emission reduction verification.

\subsection{System Architecture Overview}
\begin{figure}[ht!]
    \centering
    \includegraphics[width=\columnwidth]{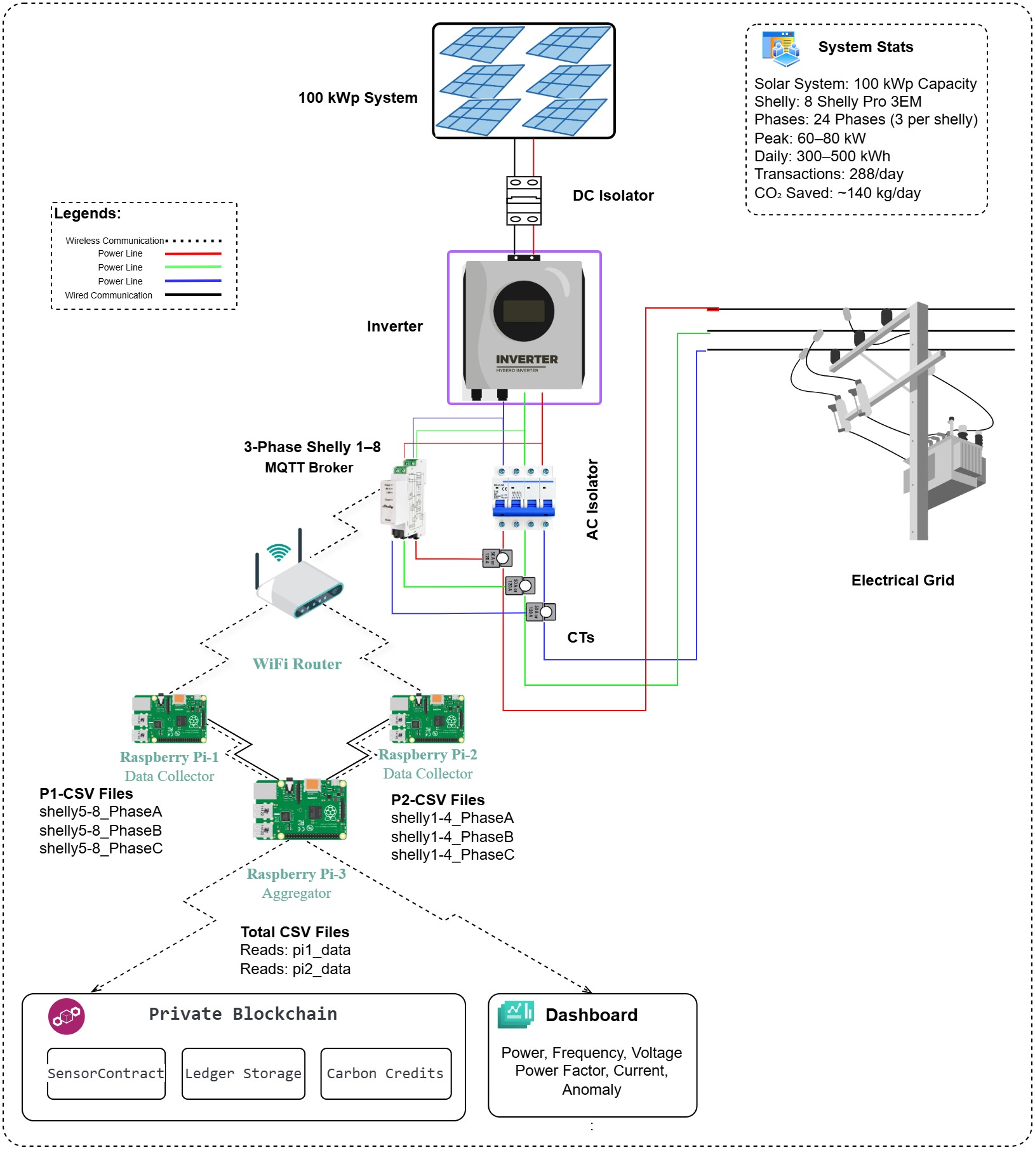}
    \caption{Framework of the proposed blockchain-based carbon-credit certification system, illustrating the flow from IoT energy measurement to on-chain verification.}
    \label{fig:Workflow}
\end{figure}

In this methodology, we propose an architectural design for a blockchain-based system to certify carbon credits, using a 100 kWp solar photovoltaic system as a reference case. The approach details how real-time power generation data can be monitored, processed, and verified. The architecture comprises four primary layers: (1) \textit{Physical Sensing Layer}, (2) \textit{Data Collection Layer}, (3) \textit{Data Aggregation Layer}, and (4) \textit{Blockchain and Certification Layer}, as shown in Figure~\ref{fig:Workflow}.

\subsubsection{Physical Sensing Layer}
The proposed \textit{Physical Sensing Layer} defines a conceptual arrangement of distributed energy meters capable of monitoring multiple electrical phase parameters within a photovoltaic inverter installation system. For illustration in Figure \ref{fig:Workflow}, the architectural proposal considers eight three-phase smart energy meters (SEM) covering a total of twenty-four phases. These meters are assumed to record parameters such as active power, voltage, current, power factor, frequency, and apparent power at intervals of one to two seconds. The commonly reported Shelly smartmeter accuracy of approximately 1\% refers to typical device data sheet specifications rather than validated performance within this study.
In the proposed framework, each smart meter is conceptually designed to transmit data over WiFi using a secure protocol with a Quality of Service (QoS), which guarantees message delivery. The sensing infrastructure is organized so that SEM communicate with a designated edge device.

\subsubsection{Data Collection Layer}

The \textit{Data Collection Layer} includes two edge-computing units, named Raspberry Pi\_A and Raspberry Pi\_B, serving as \textit{Data Collector\_A} and \textit{Data Collector\_B}. In this setup, meters 1--4 connect to Raspberry Pi\_A, while meters 5--8 connect to Raspberry Pi\_B. This structure is based on a typical photovoltaic system architecture. Both devices, shown here as Raspberry Pi 4B units, act as \textit{Data Collectors}, each tasked with receiving measurement data from the sensing layer.

The total power measured by a collector at time $t$ is expressed conceptually as:
\[
P_{\text{collector}}(t) = \sum_{i=1}^{8} P_i(t),
\qquad i \in \{\text{SEM}_1, \ldots, \text{SEM}_8\}.
\]

Each collector records measurements following a fixed schema that includes timestamps, electrical parameters, and device identifiers. Typically, measurements are stored in CSV files at a minute or as per required aligned intervals to maintain consistent timing across all phases. Several CSV files might be created each day, organized by device or phase. These collector directories are accessible via LAN, allowing the aggregation layer to access them.

\subsubsection{Data Aggregation Layer}

The \textit{Data Aggregation Layer} is defined as a third edge-computing node responsible for combining and processing the measurement files produced by the collector nodes. Through their respective directories, the aggregator will read newly generated files and compute plant-wide aggregated values.
The total plant power and the average electrical parameters at each minute are given by:
\[
P_{\text{total}}(t) = \sum_{i=1}^{24} P_i(t), \qquad 
V_{\text{avg}}(t) = \frac{1}{n} \sum_{i=1}^{n} V_i(t).
\]

For demonstration, the framework assumes that the total measurements of all phases are 34,560 records per day and is reduced to minute-level aggregates (1,440 records per day). These numerical values serve as illustrative design parameters that show data-volume reduction strategies for downstream carbon-credit certification processes.

\subsubsection{Blockchain and Certification Layer}
The \textit{Blockchain and Certification Layer} is conceptualized using a permissioned blockchain platform. The rationale for selecting a permissioned over public blockchain, addressing certification requirements, cost predictability, privacy compliance, and performance, is detailed in the platform selection discussion below.

To optimize transaction throughput and reduce operational costs, the framework groups minute-level aggregates into five-minute batches. This reduces the number of daily blockchain submissions from an illustrative 1,440 to 288. These values function solely as design assumptions to demonstrate batching strategies.

Within the system, grouped records are intended to be serialized into JSON format and sent to a designated smart contract through a dedicated channel. The smart contract conceptually validates timestamps, parameter ranges, and data structures before committing the records to the blockchain. The immutability of the ledger ensures that all energy-production entries remain verifiable for audit and certification purposes.

Within the framework, a chaincode component is designed to compute energy production and associated carbon offsets according to:
\[
E_{\text{kWh}} = \frac{P_{\text{watts}} \times t_{\text{min}}}{60 \times 1000}, 
\qquad
\text{CO}_{2,\text{kg}} = E_{\text{kWh}} \times 0.4.
\]

The architectural design comprises conceptual mechanisms and outlier detection processes. Anomalous readings are flagged and stored for audit reviews, while valid measurements contribute to the calculation of carbon credit values recorded on-chain.

\subsubsection{Blockchain Platform Selection}
The selection of Hyperledger Fabric over public blockchains is due to its compliance with carbon credit certification requirements. Compared to public alternatives, Fabric offers: (1) near-zero transaction costs versus €500-2,500 annually for Ethereum, (2) high data privacy through private channels ensuring GDPR compliance, (3) superior performance with 3,500 TPS and 2-3s latency versus Ethereum's 30 TPS and 20s latency, and (4) low energy consumption (0.5 kWh) aligning with sustainability objectives. Table \ref{tab:blockchain_comparison} compares key platforms \cite{BlockchainIoT}. Hyperledger Fabric sacrifices public transparency and full decentralization, certification applications prioritize cost predictability, regulatory compliance, and accountable validation.

\begin{table}[h]
\centering
\caption{Blockchain Platform Comparison}
\label{tab:blockchain_comparison}
\resizebox{\columnwidth}{!}{%
\begin{tabular}{lccccc}
\hline
\textbf{Platform} & \textbf{Cost} & \textbf{TPS} & \textbf{Latency} & \textbf{Privacy} & \textbf{Energy} \\
\hline

Ethereum & High & 30 & 20s & Low & Medium \\
Polygon & Medium & 7,000 & 2s & Low & Low \\
\textbf{Fabric} & \textbf{Near-zero} & \textbf{3,500} & \textbf{2-3s} & \textbf{High} & \textbf{0.5 kWh} \\
\hline
\end{tabular}%
}
\end{table}

The permissioned nature of Hyperledger Fabric requires participants to be authorized within the network. In the context of carbon credit certification, the network participants are the renewable energy producer, the certification body, and authorized auditors, a well-defined set of actors that aligns with the certification workflow described in Section~\ref{impl}. This differs from the voluntary carbon market itself, where buyers and sellers are not predefined. Our framework addresses this distinction by focusing on the certification phase: once credits are certified and recorded on-chain, the resulting certificates can be listed and traded on external voluntary market platforms (Verra, Gold Standard, etc.) where market participants operate freely. The permissioned blockchain thus serves as the certification infrastructure, not the trading marketplace, ensuring that only verified entities can issue credits while allowing those credits to circulate in open markets.

\section{Implication for Practitioners} \label{impl}

The practical deployment of blockchain-based carbon credit certification systems requires careful consideration of the regulatory landscape, certification pathways, and operational constraints. This section addresses the fundamental question practitioners face: \textit{how can carbon credits generated from renewable energy systems be utilized?}

\subsection{Regulatory Context and Market Access}

Current European legislation establishes distinct pathways for carbon credit generation and trading. Understanding these pathways is essential for determining how credits from photovoltaic systems can be used.

\subsubsection{Exclusion from Regulated Markets}

Carbon credits generated from photovoltaic energy production do not qualify for regulated carbon markets under current EU legislation. The EU Emissions Trading System (EU ETS), established by Directive 2003/87/EC \cite{directive2003ETS}, applies exclusively to high-emission industries and large-scale energy production facilities. Small and medium-scale renewable energy installations are outside its scope.

Similarly, Regulation (EU) 2018/842 \cite{reg2018effort} (Effort Sharing Regulation) sets binding emission reduction targets for Member States across sectors including agriculture and small-scale energy, but does not create individual trading mechanisms for operators. The recently adopted Regulation (EU) 2024/3012 \cite{reg2024carbon}, which establishes an EU certification framework for carbon removals, explicitly covers carbon farming, permanent carbon storage, and carbon storage in products. Technological emission avoidance through renewable energy generation is not included in this framework.

\subsubsection{Voluntary Carbon Markets as the Viable Pathway}

Given these regulatory constraints, carbon credits from photovoltaic systems can be utilized through voluntary carbon market (VCMs). These markets allow entities to voluntarily purchase carbon credits to offset their emissions, without regulatory obligation.

Voluntary markets operate through private certification standards. The dominant standards include Verra (Verified Carbon Standard), which represents approximately 66\% of voluntary carbon market volume, Gold Standard, which emphasizes projects with social and environmental co-benefits, Climate Action Reserve, which primarily operates in North America, and American Carbon Registry, which focuses on specific project types and geographies \cite{battocletti2023voluntary}.

\subsection{Certification Requirements}

To access voluntary carbon markets, practitioners must meet specific certification requirements. While each standard has particular methodologies, common requirements span five key dimensions \cite{schneider2014double}. First, \textbf{additionality} demands that the project demonstrate emission reductions would not have occurred without the project intervention. Second, \textbf{measurability} requires that emission reductions be quantifiable using accepted methodologies. Third, \textbf{permanence} ensures that reductions are maintained over time. Fourth, \textbf{verification} mandates that independent third parties verify claimed reductions. Finally, \textbf{uniqueness} ensures each credit represents a distinct emission reduction, preventing double-counting.

The blockchain architecture proposed in this work directly supports several of these requirements. The immutable ledger provides verifiable proof of energy generation data, the timestamped records enable third-party verification, and the distributed nature of the system helps prevent double-counting.

\subsection{Operational Pathway}

The process for carbon credit utilization begins with standard selection, where operators identify which voluntary market standard is most appropriate for the project characteristics and geographic location. Following this, project registration involves submitting the project to the chosen standard, providing technical specifications and baseline calculations.

The next phase is monitoring implementation, deploying the monitoring system described in Section \ref{meth} to collect verified energy production data. This is followed by third-party verification, where accredited verification bodies validate claimed emission reductions. Upon successful verification, the operator certified carbon credits. The final stage is credit utilization, where operators face a binary choice: either sell credits on voluntary markets to generate revenue, or use credits internally for sustainability reporting.

\subsection{Internal Sustainability Reporting}

In addition to being sold in the market, carbon credits can also be utilized for internal sustainability reporting. This is particularly relevant given Directive (EU) 2022/2464 \cite{directive2022CSRD} (Corporate Sustainability Reporting Directive), which requires certain enterprises to report on environmental impacts.

However, practitioners must understand a critical constraint: \textbf{carbon credits can either be sold OR used for internal claims, but not both}. Once a credit is sold, the environmental benefit transfers to the purchaser. The original generator cannot subsequently claim the emission reduction in their own reporting \cite{directive2024green}.

Besides, Directive (EU) 2024/825 \cite{directive2024green} regulates environmental claims to prevent greenwashing. Any public communication about carbon credits must be accurate, verifiable, and not misleading.

\subsection{Blockchain's Role in Market Access}

The blockchain component of the proposed architecture provides specific advantages for accessing voluntary carbon markets. Regarding \textbf{data integrity}, immutable records reduce verification costs and increase certifier confidence. The \textbf{audit trail} capability ensures that a complete transaction history is available for third-party audits required by certification standards. For \textbf{double-counting prevention}, the distributed ledger architecture makes it technically difficult to claim the same reduction multiple times \cite{marchant2022blockchain}. Finally, \textbf{transparency} enables potential buyers to verify the authenticity of offered credits.

However, blockchain is an \textit{enabler}, not a substitute for certification. Practitioners must still engage with voluntary market standards and verification bodies. The blockchain system makes this process more efficient and reliable, but does not eliminate regulatory requirements.

\subsection{Practical Constraints}

Practitioners should be aware of several practical constraints that affect the viability of carbon credit generation. Certification costs represent a significant consideration, as third-party verification incurs expenses that may be substantial relative to the credit value for smaller installations. Market price volatility means that voluntary carbon market prices vary significantly based on project type, geography, and market conditions \cite{calel2013carbon}.

The challenge of baseline establishment arises because demonstrating additionality requires establishing what would have occurred without the project, which can be complex for renewable energy installations. Finally, regulatory evolution must be monitored, as the regulatory landscape continues to develop. Regulation (EU) 2024/3012 \cite{reg2024carbon} requires multiple implementing acts from the European Commission, potentially expanding future certification pathways.

\subsection{Recommendations}

Based on the regulatory analysis and system architecture, we offer five key recommendations for practitioners. First, \textbf{early engagement} with voluntary market standard organizations during system design ensures that monitoring requirements are properly addressed from the outset. Second, \textbf{thorough documentation} of system specifications, baseline calculations, and operational data is essential for successful certification.

Third, practitioners should seek \textbf{professional guidance} from legal and technical advisors familiar with voluntary carbon markets and agricultural renewable energy. Fourth, maintaining \textbf{realistic expectations} requires calculating potential credit revenue against certification costs before committing to the certification process. Finally, operators must \textbf{choose one path}, deciding whether credits will be sold or used internally for reporting, understanding that both options are mutually exclusive.

The blockchain-based architecture presented in this work provides the technical foundation for reliable carbon credit certification, but market access ultimately depends on compliance with voluntary market standards and successful third-party verification.

\section{Threats to Validity}\label{limit}

This section discusses potential threats to validity that may affect the generalizability and applicability of the proposed framework. We acknowledge these limitations while recognizing that the work provides a concrete architectural foundation for carbon credit certification systems.

\subsection{Construct Validity}

The framework assumes smart energy meters provide approximately 1\% measurement accuracy based on typical device data sheet specifications. Although this level of precision has not been validated within our specific deployment context, it represents industry standard performance for Real-world conditions, such as internal temperature variations or electromagnetic interference, and the switching process may affect precision, and the aggregation process from 24 phases to three to five-minute batches may introduce additional measurement uncertainty. Consequently, factors suggest that carbon credit calculations should incorporate appropriate error margins during third-party verification.

The system uses a carbon conversion factor ranging from 0.25 to 1.06 kg CO$_2$ per kWh. This serves as a realistic placeholder that reflects the variation seen in European grid emission factors. However, this simplified representation does not account for the daily fluctuations in carbon intensity, which occur due to changes in the energy mix. While incorporating dynamic grid-intensity data could improve accuracy, we chose this approach to prioritize architectural clarity and system simplicity. This method aligns with common practices in voluntary carbon market methodologies, where fixed or average emission factors are frequently used for renewable-energy projects.

While the implementation details and measurement parameters are specific to photovoltaic energy production (electrical power, voltage, current), the four-layer architectural pattern, physical sensing, data collection, aggregation, and blockchain certification, is designed to be applicable to other carbon credit generation scenarios, such as forestry management, methane capture, or carbon sequestration projects, where domain-specific sensors would replace the electrical meters.

\subsection{Internal Validity}

The system relies on a stable internet connection with sufficient QoS for data transfer. Since this work mainly describes an architectural framework rather than a live system, potential network issues are discussed in theory. Such disruptions could cause data gaps, but these are operational concerns usually addressed in production using standard IoT practices like local buffering and retries—details outside this framework's scope.

Raspberry Pi devices are used as example edge computing nodes. Although they have fewer resources compared to server hardware, they are capable of handling the lightweight aggregation tasks anticipated in our design and are common in similar monitoring setups. As the framework isn't deployed yet, scalability beyond eight meters hasn't been tested; however, the modular design allows future expansion through additional collector nodes.

Blockchain performance is also considered at a high level. The batching method—reducing 1,440 raw measurements to 288 blockchain transactions daily—is a design decision to show how the framework handles data load. It is not a validated performance metric. Real-world performance will depend on specific deployment details and will need future empirical testing.

\subsection{External Validity}

The framework uses a 100 kWp photovoltaic system as an example to represent a mid-scale agricultural installation. This example aims to show how the architectural components work together, not to generalize for all system sizes. Because the framework is modular, smaller setups could use fewer meters, and larger systems could expand by replicating the collector–aggregator pattern. While certification costs might be high for very small installations, this is more a market-economic concern than a limitation of the architecture.

The selected technologies—Raspberry Pi devices, WiFi, and a blockchain backend—are shown as practical, affordable options suitable for agricultural use. These are illustrative choices, not strict requirements. Other communication methods or blockchain platforms could be used without changing the four-layer architecture. This work’s contribution is defining that pattern and demonstrating its relevance to carbon-credit certification processes, not specifying particular technologies.

The regulatory focus is on European laws, reflecting their maturity and influence on carbon-credit methods. Although certification rules differ by region, the architectural principles are broadly applicable. This geographic emphasis highlights current regulatory development levels rather than a limitation of the technical framework.

\subsection{Conclusion Validity}

The framework is conceptual and lacks empirical deployment data. This is acknowledged as a next step rather than a fundamental weakness. The work's contribution is the software architecture and its alignment with regulatory requirements, providing indications for implementation. The claimed benefits of blockchain (data integrity, audit trails, double-counting prevention) are well-established properties in the literature \cite{marchant2022blockchain}.

We do not provide detailed cost-benefit analysis or comparative evaluation against alternative approaches. This reflects our focus on demonstrating feasibility and regulatory alignment for blockchain-based certification. Economic viability will vary by context and scale, and comparative studies would require empirical implementations of multiple approaches, which is beyond the scope of this architectural work. However, the blockchain approach addresses specific trust and transparency requirements identified by certification bodies \cite{battocletti2023voluntary}, suggesting potential advantages over purely centralized alternatives for this use case.

The architectural contribution lies in integrating IoT monitoring, edge processing, and blockchain certification into a coherent system aligned with carbon credit regulatory requirements. While empirical validation remains future work, the framework provides practitioners and researchers with a concrete starting point grounded in both technical feasibility and legal compliance.

\section{Conclusions} \label{conclusion}

This paper presented a software-engineering framework for the certification of carbon credits using blockchain and IoT technologies. Building on a four-layer architecture—spanning physical sensing, data collection, aggregation, and blockchain-based certification—we formalized the core components and data flows required to support traceable and auditable carbon-credit generation. Unlike existing work that primarily targets trading mechanisms or conceptual monitoring platforms, our contribution focuses on the software structures, validation logic, and operational constraints involved in certification-oriented systems.

For practitioners, the framework offers a structured approach to designing certification pipelines that integrate heterogeneous IoT data with permissioned blockchain infrastructures. The layered design supports modularity, facilitates verification of energy-production records, and provides an auditable trail aligned with requirements of voluntary carbon-market standards. By clarifying how timestamp validation, batching strategies, and immutability can be incorporated into the architecture, the framework helps practitioners anticipate regulatory, operational, and data-quality considerations before engaging with certification bodies.

This work remains conceptual and does not include a full deployment or empirical performance evaluation. As future work, we plan to implement the proposed architecture in a real photovoltaic installation, assess system behavior under operational conditions, and quantitatively evaluate data integrity, reliability, and transaction performance. These empirical results will help validate the framework, refine its components, and provide further evidence of its applicability for certification processes in real-world settings.

Beyond empirical validation, future evolution of the framework could explore hybrid blockchain architectures that combine the compliance and privacy advantages of permissioned networks with the transparency benefits of public ledgers. One promising direction is anchoring certification records from the Hyperledger Fabric network onto a public blockchain, creating a verifiable proof of certification without exposing sensitive operational data. This approach would enable a broader blockchain-based information ecosystem where certification authenticity can be publicly verified by any market participant, while detailed production data remains within the permissioned network. Such hybrid architectures could facilitate integration with decentralized carbon credit marketplaces and support cross-platform verification mechanisms, expanding the framework's applicability beyond consortium-based certification scenarios.

\balance
\bibliographystyle{IEEEtran}

\bibliography{biblio}

\end{document}